\documentclass[10pt,twocolumn]{article}

\usepackage{multicol}
\usepackage{graphicx}
\usepackage{amsmath}
\usepackage{lipsum}
\usepackage{color}
\usepackage{siunitx}
\usepackage{authblk}
\usepackage{subfigure}
\usepackage[margin=2cm]{geometry}

\title{Drop impact on a flexible fiber}
\author[1]{Emilie Dressaire$^*$}
\author[2]{Alban Sauret}
\author[3]{Fran\c{c}ois Boulogne}
\author[3]{Howard A. Stone}
\affil[1]{Department of Mechanical and Aerospace Engineering, New York University Polytechnic School of Engineering, Brooklyn, NY 11201, USA. ( $^*$corresponding author: dressaire@nyu.edu)}
\affil[2]{Surface du Verre et Interfaces, UMR 125, 93303 Aubervilliers, France.}
\affil[3]{Department of Mechanical and Aerospace Engineering, Princeton University, Princeton, NJ 08544, USA.}
\date{ }
\begin{document}

\twocolumn[
    \begin{@twocolumnfalse}
        \maketitle
        \begin{abstract}
When droplets impact fibrous media, the liquid can be captured by the fibers or contact then break away. Previous studies have shown that the efficiency of drop capture by a rigid fiber depends on the impact velocity and defined a threshold velocity below which the drop is captured. However, it is necessary to consider the coupling of elastic and capillary effects to achieve a greater understanding of the capture process for soft substrates. Here, we study experimentally the dynamics of a single drop impacting on a thin flexible fiber. Our results demonstrate that the threshold capture velocity depends on the flexibility of fibers in a non-monotonic way. We conclude that tuning the mechanical properties of fibers can optimize the efficiency of droplet capture. \\
            \medskip\medskip\medskip
        \end{abstract}
    \end{@twocolumnfalse}
]

\section{Introduction}

Drop impacts on a solid substrate result in short-time dynamics that involve spreading, splashing, receding and/or bouncing \cite{clanet2004,yarin2006,Deegan2008,eggers2010,bird2010}. Understanding and controlling the post-impact dynamics, including the interaction with the substrate \cite{Bird2013, Lhuissier2013}, the capture of the drop and its splashing, i.e. the detachment of small satellite droplets, is critical to the efficiency of various engineering processes. For industrial spray coating, ink-jet printing and pesticide delivery, drop capture is necessary and splashing is to be avoided \cite{Bergeron2000}, whereas for fuel combustion, the ejection of small droplets is essential \cite{Eggers2008}.

Previous studies of drop impact have highlighted the importance of the substrate on the dynamics of the impact and capture of the drops \cite{xu2005,Mock2005,Xu2007}. The surface characteristics of the substrate including its roughness and chemistry are known to modify the splash behavior \cite{Stow1981,Tsai2010}. Also, the spreading of the liquid can be enhanced through forced imbibition of the surface roughness \cite{Range1998,lorenceau2003,delbos2010}.

The mechanical properties of the substrate and its ability to move following the impact \cite{Dickerson2012,Dickerson2014} influence the post-impact dynamics: for instance soft substrates such as clamped elastic sheets can be used to suppress splashing \cite{Pepper2008}. In addition, the impact of a drop on a beam clamped at one end has been investigated \cite{soto2014,Gart2015} as the role of substrate compliance is relevant to agricultural and industrial applications such as the treatment of  leaves with pesticides or herbicides \cite{williamson1983,lightbody1985}, or the spray cooling and coating of flexible structures.

Finally, the geometrical features of the substrate can strongly influence the behavior of the drop during and after impact. Indeed, for fiber-like substrates whose typical size are comparable to the diameter of the impacting drop, then drops can either be captured or pass through the substrate. Hung and Yao \cite{hung1999} studied the impact dynamics on horizontal rigid fibers and later Lorenceau et al. \cite{lorenceau2004} showed that the ability of a fiber to capture a drop depends on the drop radius and the impact velocity: above critical values of these two parameters, the drop is released upon impact. The threshold velocity above which the drop is no longer captured has also been shown to depend on the inclination of the fiber \cite{piroird2009} and the position of the impact with respect to the fiber \cite{lorenceau2009}. When a drop is captured, its spreading or absence thereof leads to different equilibrium morphologies that have been characterized extensively by studies on individual or pairs of fibers \cite{huang2009,gilet2009,gilet2010,protiere2013,song2013,sauret2014,sauret2015,Weyer2015, sauret2015b}. In this wetting scenario, the elastic deformation of the fibers leads to spreading dynamics that differ from the rigid case \cite{bico2004,duprat2012}.

The role of flexibility in drop capture by fiber-like structures is critical because of their ubiquitous presence in natural and industrial systems such as the stems of plants, needles and narrow leaves, fiber filters and glass wool.  To the best of our knowledge, no study has thus far characterized the dynamics of drop impact and capture on a flexible thin fiber. We report a comprehensive experimental study on the effects of flexibility on the capture threshold using a fiber clamped at one end. We find that the threshold velocity for capture is strongly affected by the compliance of the fiber: as the length of the fiber increases, the threshold velocity for capture first increases monotonically and then reaches a maximum, suggesting an optimum flexibility for drop capture. We present a model for the threshold velocity as a function of the flexibility of the fiber, which allows us to rationalize our experimental results by comparing the typical time scales for the drop to pass through the fiber and for the fiber to bend.

\section{Experimental setup and observations}

\begin{figure}[h!]
    \begin{center}
        \subfigure[]{\includegraphics[width=0.48\textwidth]{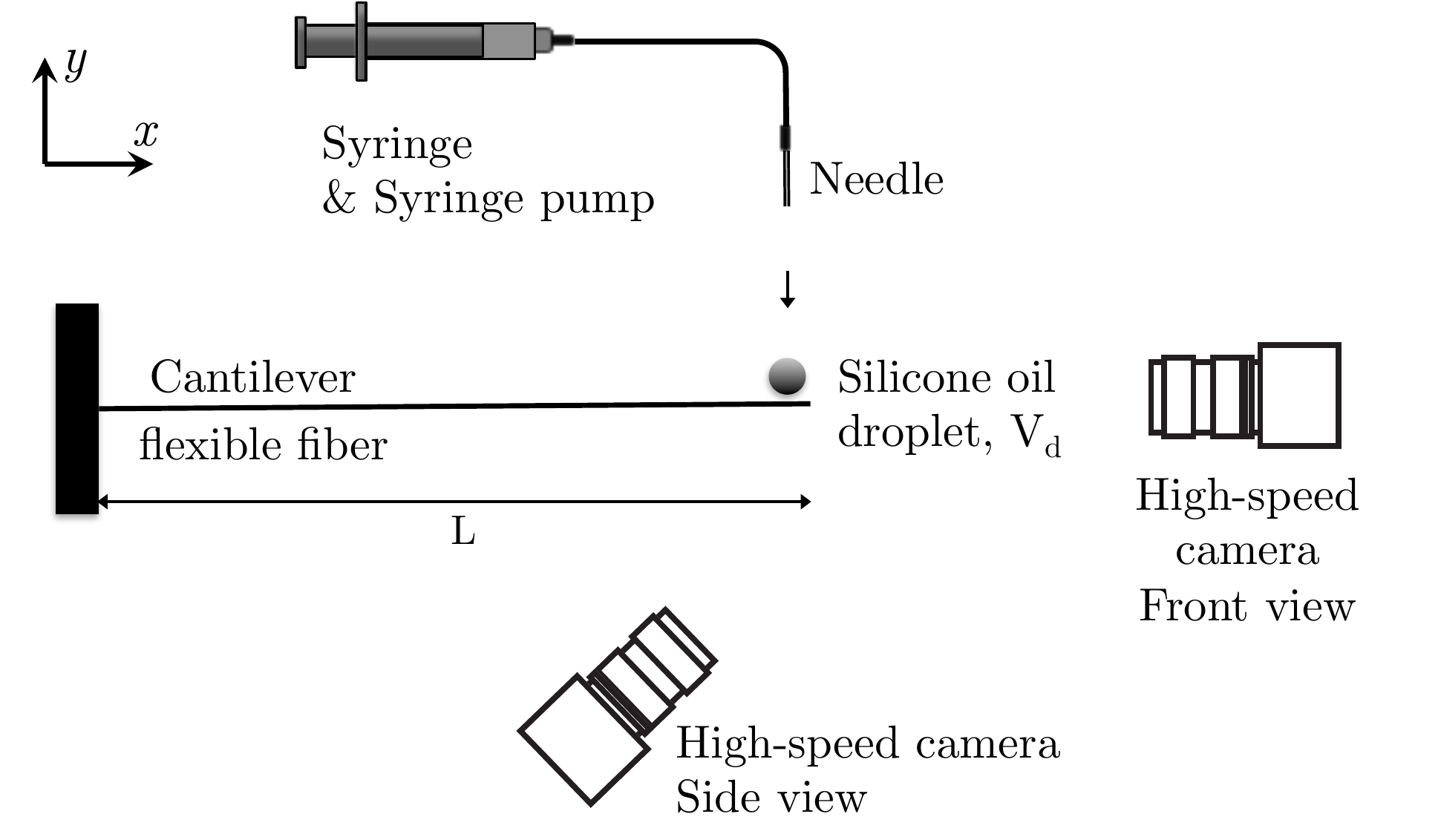}}\\
        \subfigure[]{\includegraphics[width=0.43\textwidth]{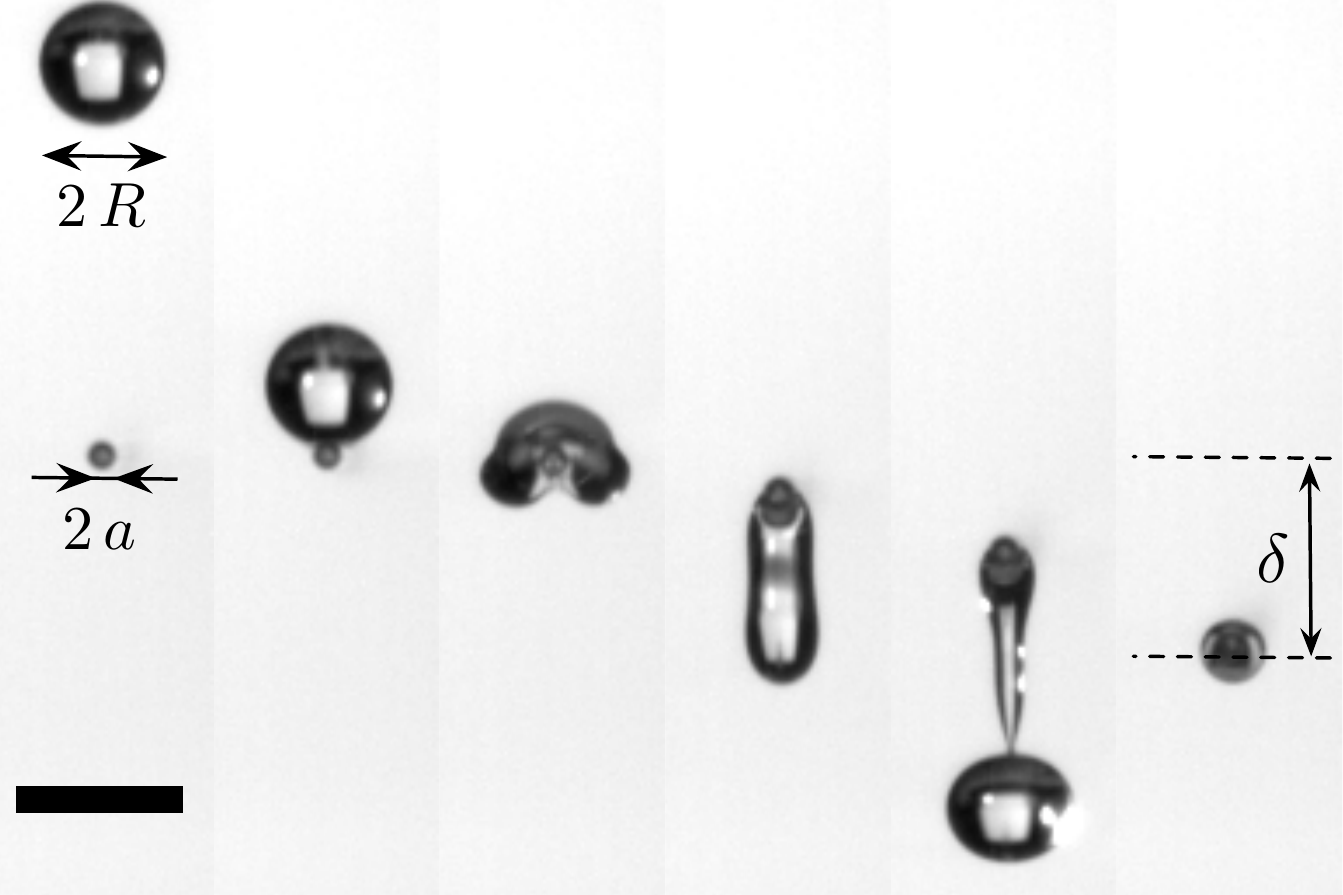}}
    \end{center}
    \caption{ \small (a) Schematic of the experimental setup. A high-speed camera perpendicular to the axis of the fiber records the motion of the drop and the oscillations of the fiber upon impact. (b) The impact of the drop of radius $R$ on the fiber of radius $a$ is also recorded from the front view using a second high-speed camera aligned with the axis of the fiber to ensure that the impact is centered. $\delta$ denotes the maximum deflection of the fiber. From left to right $t=[-9,\, 0, \, 2, \, 7, \, 16.8, \, 38.6]$ ms. Scale bar: 2 mm.}
    \label{fig:setup}
\end{figure}

To study the impact of a liquid drop on a flexible fiber of length $L$, we clamp horizontally a borosilicate glass capillary (VitroCom) at one extremity only, so that it is free to bend along its length (see Fig. \ref{fig:setup}(a)). The fiber has an outer diameter of $400\,\mu$m, a length $L\in [4;\,300]\,{\rm mm}$, a linear density $\lambda_m=1.2\,\textrm{mg.cm}^{-1}$ and we measure its bending stiffness $B=E\,I=4.34\,\times 10^{-5}\,\textrm{N.m}^2$.
To prevent the liquid from filling the capillary, the tip is plugged with epoxy resin over a length of less than $1$ mm. The small size of this plug ensures that the mechanical properties of the fiber are not significantly modified.

For the experiments presented in this paper, we use different silicone oils whose respective kinematic viscosity, density and surface tension are (i) $\nu=10^{-6}\,{\rm m^2.s^{-1}}$, $\rho = 820 \, {\rm kg.m^{-3}}$, $\gamma = 17.5\,{\rm mN.m^{-1}}$, (ii) $\nu=5\,\times 10^{-6}\,{\rm m^2.s^{-1}}$, $\rho = 920\, {\rm kg.m^{-3}}$, $\gamma = 19.7\,{\rm mN.m^{-1}}$ and (iii) $\nu =  3.5\,\times 10^{-6}\,{\rm m^2.s^{-1}}$, $\rho = 910 \,{\rm kg.m^{-3}}$, $\gamma = 19\,{\rm mN.m^{-1}}$. Droplets of silicone oil are released from different heights to tune the velocity of impact on the fiber, $V_d =100-800\,\,\textrm{mm.s}^{-1}$. Drops of different radii, from $R=0.41\,\pm\,0.02\,\textrm{mm}$ to $R=1.05\,\pm\,0.02\,\textrm{mm}$, are generated using blunt needles (20G, 27G and 30G) and glass capillaries tapered with a pipette puller.
These needles are connected to a syringe pump to slowly form a pendant drop, which subsequently falls under its own weight.
The position of impact on the fiber is controlled by clamping the fiber on a translation stage with a micrometer screw (PT1, Thorlabs), and monitored with a high-speed camera (Photron) that records the front view of the impact to determine if the drop is centered (Fig. \ref{fig:setup}). If the impact of the drop is off-centered, the threshold velocity decreases and therefore controlling the position of the impact is crucial in these experiments \cite{lorenceau2009}. This method also ensures that the motion of the fiber is in the vertical plane $(x,\,y)$. In addition, a second high-speed camera perpendicular to the axis of the fiber records the impact of the drops and the deflection of the fiber. We then use a custom-written image-analysis code to determine the impact velocity $V_{d}$ and the deflection of the fiber over time. {To study the influence of the fiber length, we perform experiments with drops of silicon oil ($\nu=10^{-6}\,{\rm m^2.s^{-1}}$ and $\nu=5\,\times10^{-6}\,{\rm m^2.s^{-1}}$) of radius $R=0.76\,{\rm mm}$. We also vary systematically the radius of drops of silicon oil ($\nu =  3.5\,\times 10^{-6}\,{\rm m^2.s^{-1}}$) for a given fiber length, $L=140\,{\rm mm}$.}

\begin{figure}
    \begin{center}
	\includegraphics[width=0.48\textwidth]{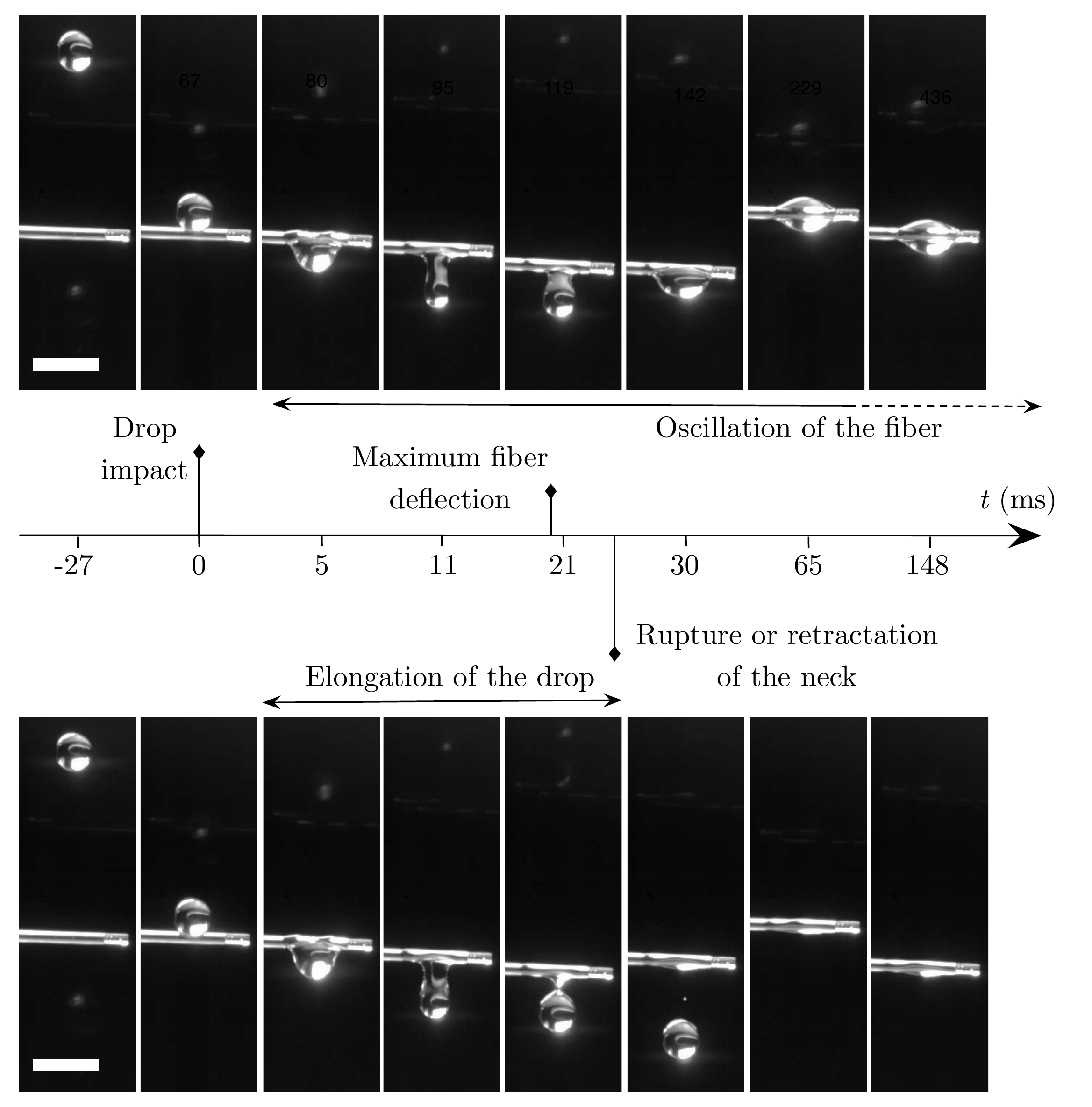}
    \end{center}
    \caption{ \small Image sequences of a drop of radius $R=0.76$ mm impacting a fiber ($L=14$ cm, $2\,a=400\,\mu$m) at different velocities $V_{d}$ (side view). For $V_{d}=370\,{\rm mm.s^{-1}}$ the drop is captured by the fiber (top images) whereas for $V_{d}=390\,{\rm mm.s^{-1}}$ the drop is released (bottom images). Scale bar: $2\,{\rm mm}$.}
    \label{fig:phenomenology}
\end{figure}

The impact of a drop on a flexible fiber triggers a series of events that depends on the impact velocity $V_d$. The two main regimes observed experimentally are reported in Fig. \ref{fig:phenomenology} where they correspond to two slightly different values of the impact velocity: $V_{d}=370$ mm.s$^{-1}$ and $V_{d}=390$ mm.s$^{-1}$.
Upon impact, the fiber deflects downward owing to its flexibility. The fiber tip and the drop travel downward together, but at different velocities. When the distance between the fiber tip and the center of mass of the drop reaches a maximum, the bridging neck of liquid can either retract in the capture regime (Fig. \ref{fig:phenomenology}, upper series of images) or break in the release regime (Fig. \ref{fig:phenomenology}, lower series of images). In both regimes, the fiber continues oscillating before reaching an equilibrium position after a time scale much longer than the time scale of the drop-fiber interaction.

When the drop impinges on the fiber at a typical speed of $100$ to $800$ mm.s$^{-1}$, gravity and inertia cause the drop to leave the fiber, whereas viscous and capillary forces lead to the capture of the drop. Different dimensionless parameters compare these different effects. The Weber number characterizes the relative influence of inertial and interfacial effects, $We={2\,\rho \,{V_d}^2\,R}/{\gamma} \simeq 2-40 $. The importance of viscous and interfacial effects is described by the capillary number $Ca=\eta\,{V_d}/\gamma \sim 0.01-0.2$. For the low viscosity liquids used in this study, the capillary force is mainly responsible for the capture of the drop on the fiber. Finally, the Reynolds number based on the drop diameter, {$Re=We/Ca=2\,V_d\,R/\nu \sim 1200-1600$} showing that inertial effects are much larger than viscous effects. Therefore, the capture dynamics will usually result from a balance between inertia and capillarity. 

\begin{center}
    \begin{figure}
\includegraphics[width=0.48\textwidth]{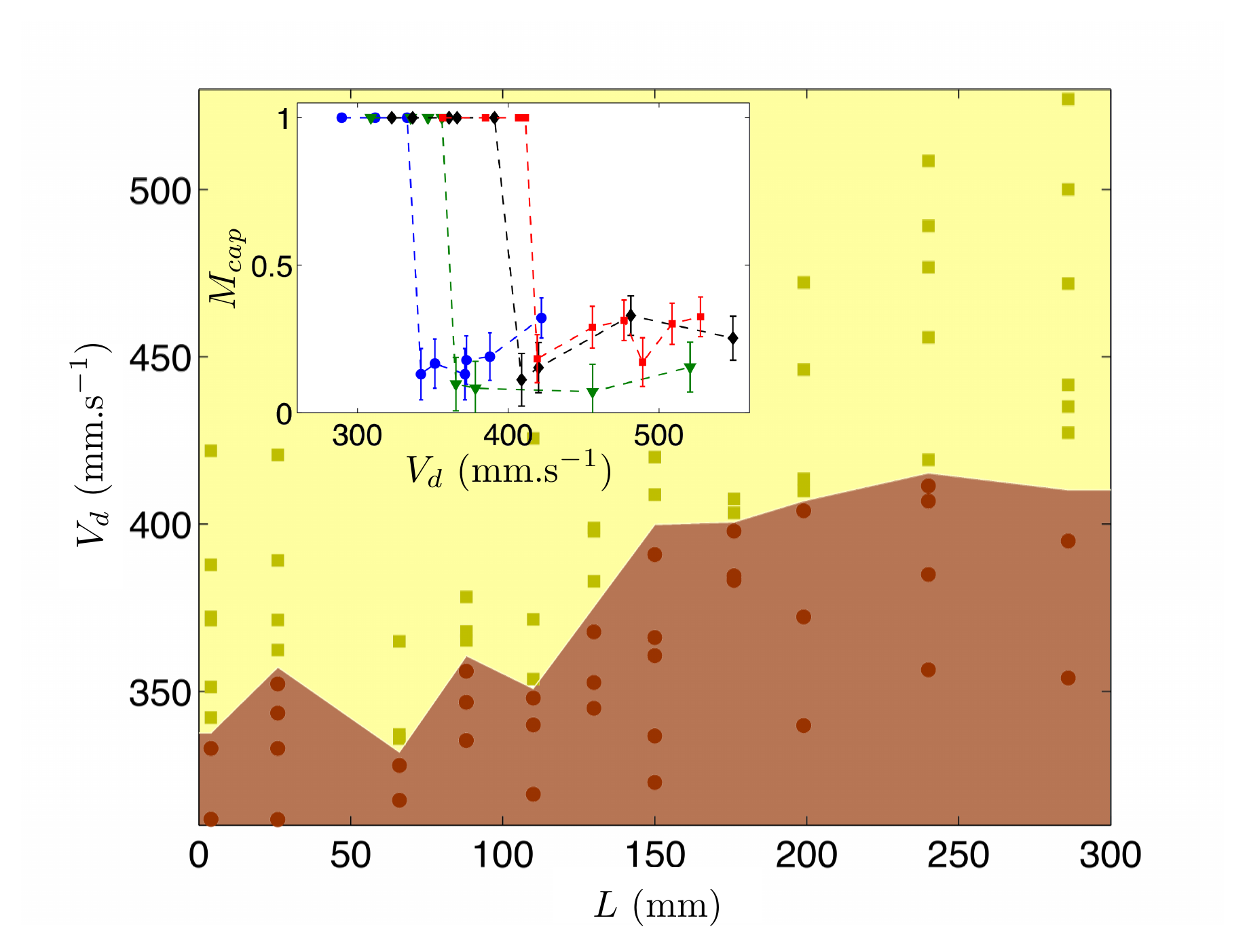}
        \caption{ \small Regime diagram and evolution of the threshold velocity $V_{d}^*$ for capture as a function of the fiber length $L$. Inset: Variation of the relative captured mass $M_{cap}=m_{cap}/m_{d}$, where $m_{cap}$ is the mass of liquid captured by the fiber, versus the drop velocity for $L= 1$ mm (blue), $L= 88$ mm (green), $L=150$ mm (black) and $L=240$ mm (red). For $M_{cap}=1$, the drop is captured, and for  $M_{cap}<1$, the drop is released. {The experiments are performed with $R=0.76\,\textrm{mm}$ and $\nu=5\,\times 10^{-6}\,{\rm m^2.s^{-1}}$}}
        \label{figure_capture}
    \end{figure}
\end{center}

To study the drop capture and drop release regimes, we systematically vary the velocity of the drop and the fiber length thus modifying the fiber flexibility. We measure the mass of liquid captured by the fiber (based on an average of results for $10$ to $20$ drops) and rescale the captured mass with the mass of the impacting drop: $M_{cap}=m_{cap}/m_{d}$ (inset of Fig. \ref{figure_capture}). {For a given length, we observe that} as the impact velocity of the drop increases, the transition from $M_{cap}=1$ (drop capture) to $M_{cap} \sim 0.1-0.4$ (drop release) defines the threshold velocity $V_d^*$. {In addition, the proportion of liquid remaining on the fiber when the drop is not entirely captured is of the order of $10\%$-$30\%$.}
As illustrated in Fig. \ref{figure_capture}, increasing the length of the fiber, hence its flexibility, affects the threshold velocity though the response appears non-monotonic. To understand these trends, we first model the response of the elastic fiber to the impact before considering the drop/fiber interaction.

\section{Fiber dynamics}

Upon drop impact, the fiber starts oscillating and its tip describes a damped harmonic oscillation as illustrated in Fig. \ref{figure_fiber}(a). Since some liquid (between 10\% and 100\%) remains on the fiber, the tip oscillates around an equilibrium position that corresponds to a static downward deflection $c$. The magnitude of the deflection $c$ is larger when the drop is captured (see Fig. \ref{figure_fiber}(a)) since the quantity of liquid that remains attached is greater.

\subsection{Vibration frequency}

\begin{figure*}
    \begin{center}
       \subfigure[]{\includegraphics[width=0.46\textwidth]{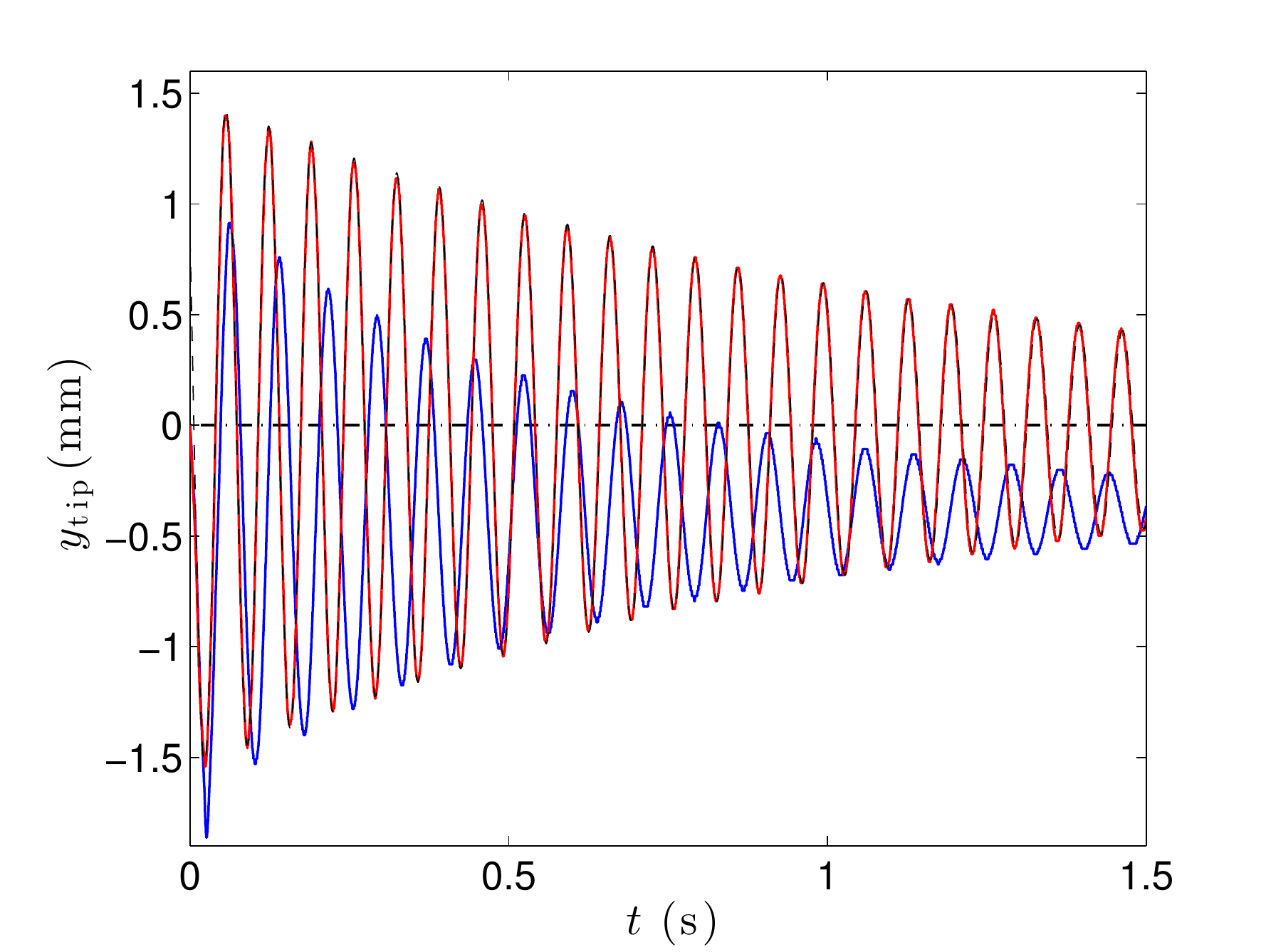}}
       \subfigure[]{\includegraphics[width=0.46\textwidth]{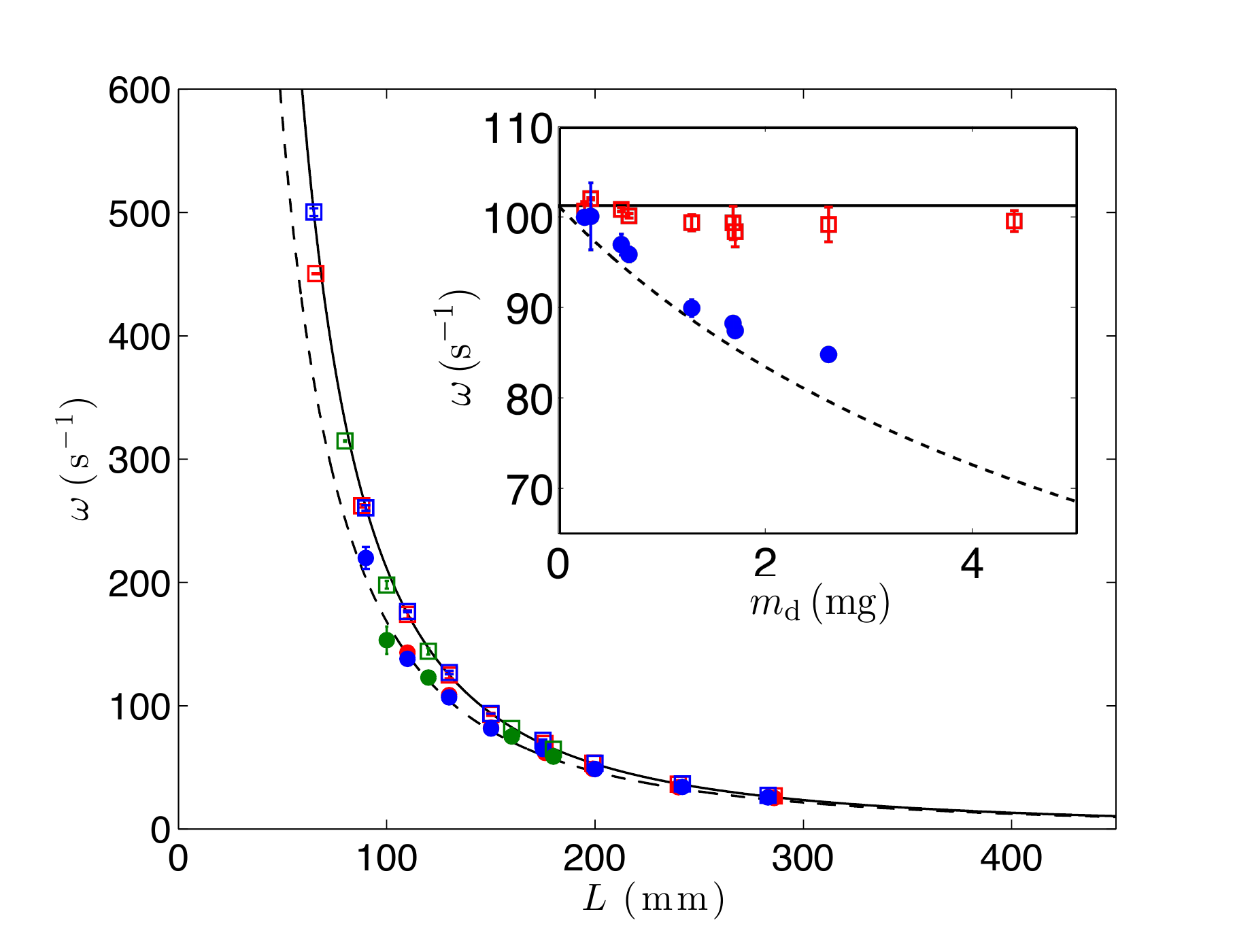}} \\
       \subfigure[]{\includegraphics[width=0.46\textwidth]{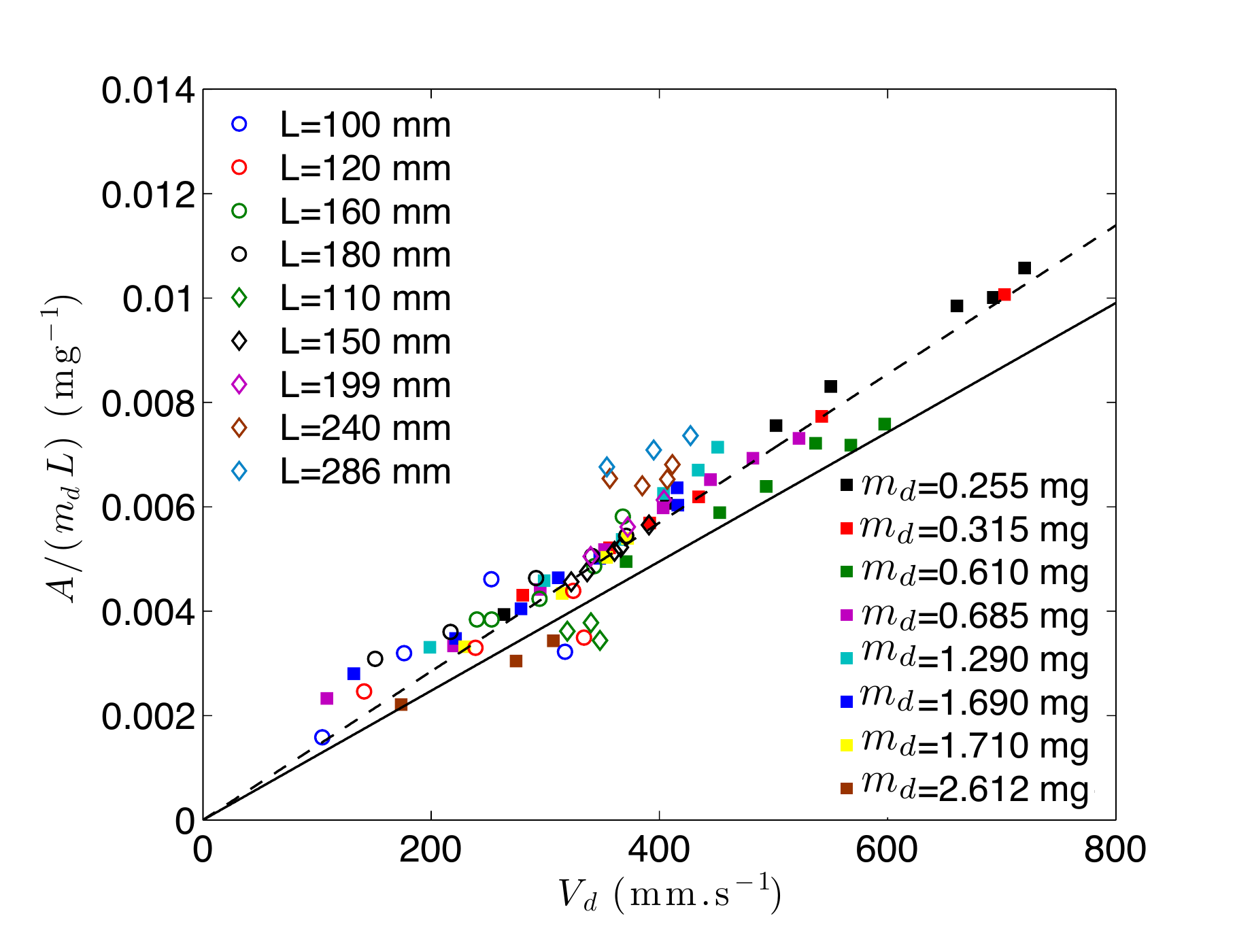}}
       \subfigure[]{\includegraphics[width=0.46\textwidth]{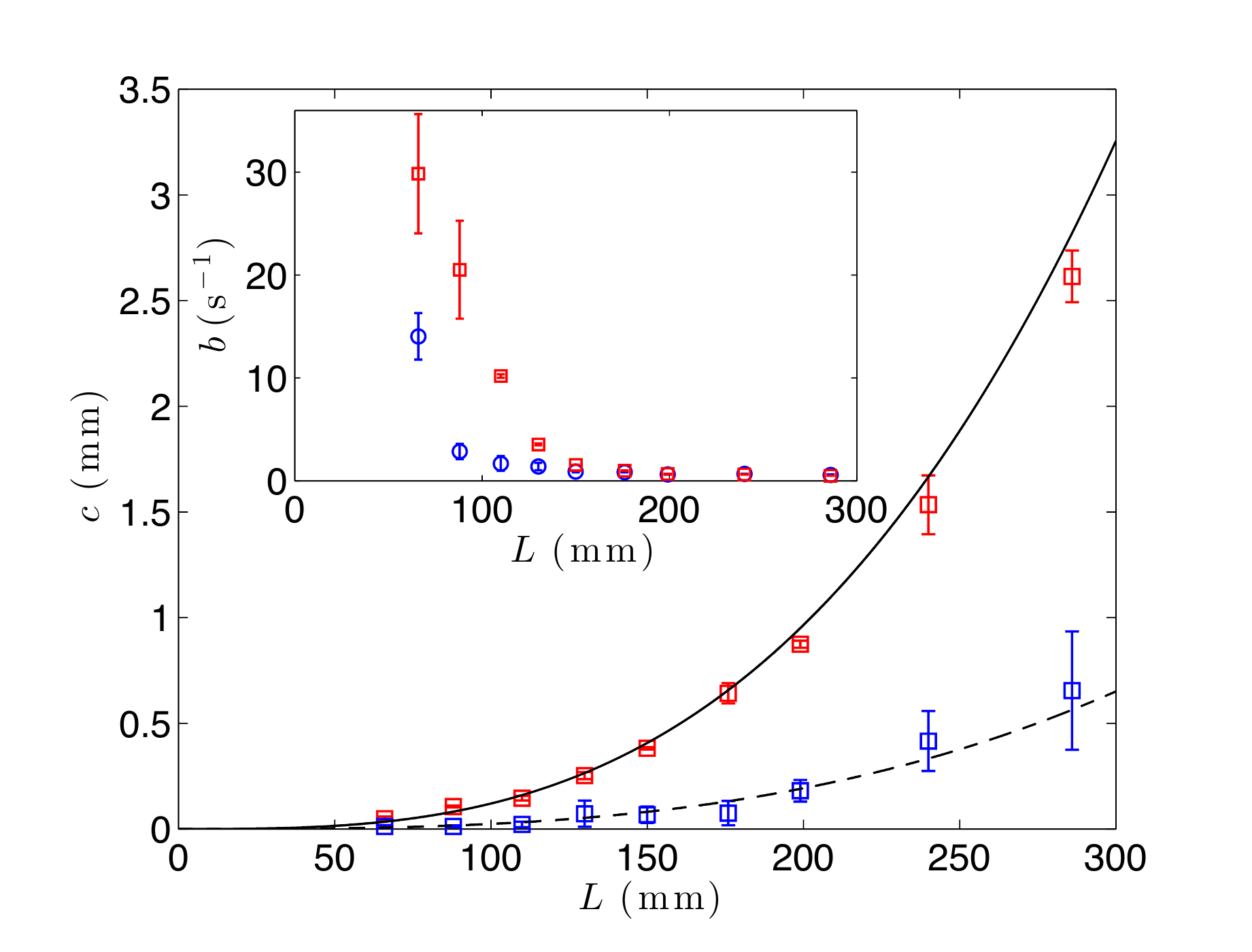}}
        \caption{ \small (a) Displacement of the tip of a $150\,{\rm mm}$-long fiber versus time when the drop is captured (in blue) or released (in red). The black dashed curve is the fit obtained using equation (\ref{eq:osc}) and the horizontal dashed-dotted line indicates the initial position of the fiber.
            (b) Vibration frequency {determined over all the oscillations} versus the length of the fiber when a drop is captured (filled symbol) or released (hollow symbol) for $5\,\times 10^{-6}\,{\rm m^2.s^{-1}}$ (red), $\times 10^{-6}\,{\rm m^2.s^{-1}}$ (blue) and $3.5\,\times 10^{-6}\,{\rm m^2.s^{-1}}$ (green) silicone oil. Inset: evolution of the vibration frequency for increasing drop masses (red: drop released; blue: drop captured), {$\nu=3.5\,\times 10^{-6}\,{\rm m^2.s^{-1}}$ and $L=140\,{\rm mm}$}. On both figures, the theoretical first-mode frequency is also shown (continuous line: drop released; dashed line: drop captured).
        (c) Evolution of the rescaled amplitude $A/(m_d\,L)$ of the fiber deflection at impact as a function of the impact velocity $V_d$ when the drop is captured. $3.5\,\times 10^{-6}\,{\rm m^2.s^{-1}}$ silicone oil (varying radius of the droplet and $L=100,\,120,\,160,\,180$ mm) and $5\,\times 10^{-6}\,{\rm m^2.s^{-1}}$ silicone oil ($L=110,\,150,\,199,\,240,\,286$ mm, constant radius $R=0.76$ mm of the droplet) are used. The continuous line is the relation (\ref{eq:amplitude_A}) and the dotted line is the best fit.
        (d) Evolution of the static deflection of the fiber tip $c$ after the oscillations versus the length of the fiber $L$; the black and dashed lines are the theoretical expression (\ref{eq:amplitude_c}) with $m_{cap}=m_d$ and $m_{cap}= 0.2 \, m_d$, respectively. Inset: damping coefficient $b$ as a function of the length of the fiber. The red squares indicate drop capture and the blue circle the drop release. {Experiments are performed with $R=0.76\,{\rm mm}$ and $\nu=5\,\times 10^{-6}\,{\rm m^2.s^{-1}}$}}
        \label{figure_fiber}
        \end{center}
    \end{figure*}

To describe the free oscillations of the flexible fiber following the impact of the drop at the tip of the fiber, we use the Euler-Bernoulli beam equation \cite{Timoshenko1959}:
\begin{equation}
    E\,I\,\frac{\partial^4 y(x,t)}{\partial x^4}=-\lambda_m\,\frac{\partial^2 y(x,t)}{\partial t^2}-b\,\frac{\partial y(x,t)}{\partial t},
    \label{Euler_Bernoulli}
\end{equation}
where $y(x,t)$ measures the transverse displacement or deflection of the fiber. In this equation, the last term describes the damping during the oscillation of the fiber and $b$ is the damping coefficient. Equation (\ref{Euler_Bernoulli}) can be solved using the appropriate boundary conditions to obtain the deflection of the fiber tip \cite{Timoshenko1959}:
\begin{equation}\label{eq:osc}
    y_{tip}(t) = A e^{-b\,t} \sin\left(\omega\,t + \phi\right) + c,
\end{equation}
where $\omega$ is the vibration frequency of the beam, $A$ the amplitude of oscillation, $\phi$ a phase, and $c$ the equilibrium position of the fiber at $t \to + \infty$. Experimentally, we observe that the fiber oscillates at the frequency of the first vibration mode. When the drop is not captured by the fiber, the analytical expression of the first mode is
\begin{equation}\label{vibration_1}
    \omega=(1.875)^2\,\left(\frac{E\,I}{\lambda_m\,L^4}\right)^{1/2},
\end{equation}
as the mass of liquid at the tip of the fiber can be neglected. When the drop is captured by the fiber, the oscillation frequency is modified by the mass of liquid $m_{d}$ at the tip of the fiber, where the drop impinges on the fiber. In this situation, the frequency of the first mode can be approximated by \cite{Timoshenko1990,Rao1995,Macho-Stadler2015}:
\begin{equation}\label{vibration_2}
    \omega' \simeq \frac{(1.875)^2}{1+4.1\,m_{d}/(\lambda_m\,L)}\,\left(\frac{E\,I}{\lambda_m\,L^4}\right)^{1/2}.
\end{equation}
The vibration frequencies measured for different impact velocities $V_d$, drop viscosities and fiber lengths are shown in Fig. \ref{figure_fiber}(b). The comparison between the expressions (\ref{vibration_1}) and (\ref{vibration_2}) and the experimental results obtained for a fiber of given length, $L=140\,{\rm mm}$, and varying drop mass, $m_{d}$, are shown in the inset of Fig. \ref{figure_fiber}(b). {We observe that the presence of an additional mass at the tip of the fiber in the drop capture regime can modify significantly the frequency of oscillation for large drop.} The analytical expressions of the vibration frequencies for both regimes compare well with our experimental findings and can be used below to describe the motion of the fiber.

\subsection{Fiber deflection}
To determine the amplitude $A$ of the oscillations of the fiber tip, we consider the momentum transfer from the impacting drop to the fiber. Before the impact, a drop of mass $m_d$ travels with a velocity $V_d$ and the fiber of mass $\lambda_m\,L$ is at rest, i.e. $V_{fib}=0$. After impact, the fiber tip and the drop travel a distance $A$ in a time equal to one quarter of a period, $\pi/(2\,\omega)$. Therefore, the momentum balance between the drop and the beam is
\begin{equation}\label{bkchouch}
m_d\,V_d \simeq A\,\left(m_d+\frac{\lambda_m\,L}{2}\right)\,\frac{2\,\omega}{\pi}.
\end{equation}

Considering the situation in which the mass of the drop is much smaller than the mass of the fiber, equation (\ref{bkchouch}) can be simplified to
\begin{equation}\label{eq:amplitude_A}
A \simeq \frac{\pi}{(1.875)^2}\,\frac{m_d\,L\,V_d}{(\lambda_m\,E\,I)^{1/2}}.
\end{equation}

Based on the form of (\ref{eq:amplitude_A}), we report the rescaled amplitude of oscillation $A/(m_d\,L)$ for varying lengths of fiber and mass of impacting drops in Fig. \ref{figure_fiber}(c). To ensure the total momentum transfer, we consider the capture regime. In the range of approximations made to obtain equation (\ref{eq:amplitude_A}), we find a fairly good agreement with the experimental results. The data show that $A \propto m_d\,L\,V$ and that the prefactor has the expected order of magnitude since the best fit is obtained by multiplying (\ref{eq:amplitude_A}) with a prefactor $\kappa \simeq 1.15$. This correction could be due to the deformation of the drop that can dissipate some energy.

In addition, when a mass of liquid $m_{cap}$ lies at the fiber tip, the fiber deflects downward under the liquid weight applied at its tip. The amplitude of the deflection $c$ can be obtained from a static torque balance on the cantilever beam \cite{Gere1984}:
\begin{equation}
    c = \frac{m_{cap}\,g\,L^3}{3\,EI} \label{eq:amplitude_c}.
\end{equation}
This expression fits well the experimental results reported in Fig. \ref{figure_fiber}(d) with $m_{cap}= m_d$ in the capture regime and $m_{cap}\sim 0.2 \, m_d$ in the release regime (see inset of Fig. \ref{figure_capture}).

Finally we estimate the damping coefficient $b$ of the oscillations for both regimes, drop capture and release, as presented in the inset of Fig. \ref{figure_fiber}(d). The damping coefficient decreases with increasing fiber length. However, because the time scale of the drop/fiber interaction, typically a few tens of milliseconds, is negligible compared to $1/b$, which is typically of the order of one second, this parameter is not relevant to evaluate the influence of the fiber deflection on drop capture and will be neglected in the rest of this study.

In summary, using the equations derived above, we are able to describe the motion of the fiber tip after the impact of the drop and can characterize its influence on the capture threshold. We next consider the interplay between the impact of the drop, its elongation and the motion of the fiber tip.\\

\section{Drop capture or release}
\subsection{Bending time versus elongation time}
\begin{center}
    \begin{figure}[h!]
\includegraphics[width=0.48\textwidth]{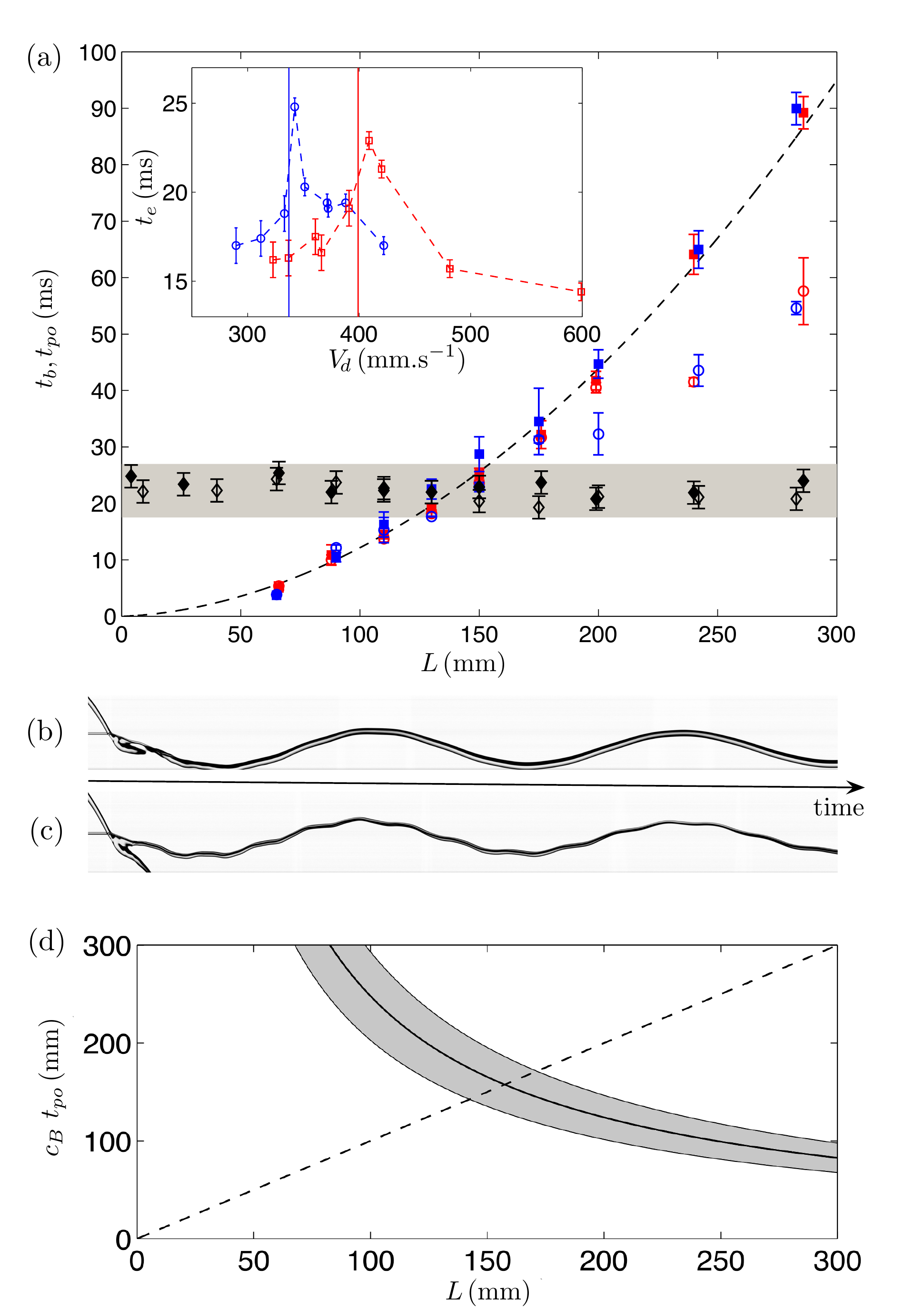}
        \caption{ \small (a) Evolution of the bending time $t_{\rm b}$ as a function of the length of the fiber for $R=0.76\,\textrm{mm}$ and $\nu=1\,\times 10^{-6}\,{\rm m^2.s^{-1}}$ (blue) or $\nu=5\,\times 10^{-6}\,{\rm m^2.s^{-1}}$ (red) for drop capture (filled) and release (hollow). The dashed line represents the relation $t_b= 1.3\,\pi/(2\,\omega)$. The black diamonds indicate the pinch-off time $t_{\rm po}$ for  $\nu=1\,\times 10^{-6}\,{\rm m^2.s^{-1}}$ (hollow) and $\nu=5\,\times 10^{-6}\,{\rm m^2.s^{-1}}$ (filled) and the grey region is the range of pinch-off times. Inset: the elongation time versus impact velocity $V_d$ for {$R=0.76\,{\rm mm}$}, $\nu=5\,\times 10^{-6}\,{\rm m^2.s^{-1}}$ silicone oil and $L=4$ mm (blue) and $L=150$ mm (red). The vertical lines indicate the capture threshold determined experimentally. {The elongation time obtained at the threshold defines the pinch-off time, i.e. the time when the drop is in contact with the fiber $t_{po}$}. (b) and (c) are spatio-temporal diagrams showing the first oscillations of the fibers following the impact for drop capture and release, respectively. (d) Analytical comparison of the distance travelled by the bending wave during the time when the drop is in contact with the fiber, $t_{po}=22\,\pm 4\,{\rm ms}$ versus the length of the fiber. The grey region represents the uncertainties on the travelled length due to the uncertainties on the pinch-off time. The dashed line corresponds to $y=x$.}
        \label{figure_5}
    \end{figure}
\end{center}

\vspace{-0.7cm}
After the impact, two phenomena occur simultaneously while the drop and the fiber travel together, as illustrated in the time line in Fig. \ref{fig:phenomenology}: the drop crosses the fiber while the fiber bends as described above. If the characteristic time scale, $t_{\rm b}$, for the fiber tip to reach its lowest position is much larger than the characteristic time scale for the drop to cross the fiber, only limited displacement of the fiber tip occurs during the interaction. More significantly, when the crossing time and the bending time are comparable, the displacement of the fiber tip is not negligible and affects the drop dynamics.

We define the elongation time of the drop $t_{e}$ as the time between the instant when drop first touches the fiber and the instant at which the narrow liquid bridge that connects the drop to the fiber stops stretching. At this point, the liquid neck can either break (in the release regime) or widen as the drop recoils toward the fiber (capture regime). We report the values of $t_{e}$ measured for two viscosities, $\nu=10^{-6}\,{\rm m^2.s^{-1}}$ and $\nu=5\,\times 10^{-6}\,{\rm m^2.s^{-1}}$, and different fiber lengths (see Fig. \ref{figure_5}(a)). For a given length, we observe that the elongation time first increases until the threshold between the capture and release regimes (see inset of Fig. \ref{figure_5}(a)). In the drop release regime, the elongation time decreases as the velocity at the impact increases. In what follow, we refer to the maximum value of $t_e$, corresponding to the transition, as the pinch-off time, noted $t_{po}$ {as can be observed in the inset of Fig. \ref{figure_5}(a)}. We find that the pinch-off time lies in the range $t_{po}=22 \pm 4\,{\rm ms}$ for the drop size used in this set of experiments, $R=0.76$ mm. Note that the pinch-off time depends slightly on the drop size as illustrated in Appendix A. However, for our systematic experiments varying the fiber length and using two different viscosities, the drop size remains constant and the pinch-off time is about $22$ ms.

Then, to compare quantitatively the time scale associated with the motion of the fiber tip and with the elongation of the drop, we measure the bending time of the fiber $t_b$ for varying lengths when the drop is captured or released and report results for two different viscosities, $\nu=10^{-6}\,{\rm m^2.s^{-1}}$ and $\nu=5 \,\times 10^{-6}\,{\rm m^2.s^{-1}}$ (Fig. \ref{figure_5}(a)). The bending time corresponds approximately to the time required for the fiber to move from its initial position to its maximum deflection. Considering the experimental fit defined by the equation (\ref{eq:osc}), $y_{tip}(t)=A\,{\rm e}^{-b\,t}\,\sin(\omega\,t+\phi)+c$, the bending time is $t_b= \pi/(2\,\omega)$.

Experimentally, we observe that $t_b$ is indeed proportiona l to $\pi/(2\,\omega)$ with a fitting prefactor $\alpha \simeq 1.3$ (Fig. \ref{figure_5}(a)).  {This small difference could be due to two effects. First, the fiber needs to travel over a distance $\delta=A+c$ and the assumption $t_b= \pi/(2\,\omega)$ is based only on the travel over a distance $A$. Second, the drop undergoes large deformations during the impact which affects the impact dynamics}. We also observe that for long fibers (typically $L > 150\,{\rm mm}$), the measured bending time becomes different in the two regimes (drop capture represented with solid symbols and drop release represented with hollow symbols in Fig. \ref{figure_5}(a)). The spatio-temporal diagrams for drop capture (Fig. \ref{figure_5}(b)) and drop release (Fig. \ref{figure_5}(c)) show that the fiber oscillation exhibits two frequencies when the drop is released: the first and secondary modes of the fiber. Therefore the measured maximum of deflection $\delta=A+c$ can be modified.

To understand the presence of the secondary oscillation, we need to consider that the impact of the drop generates a bending wave that travels along the axis of the fiber at a speed \cite{Love1944}
\begin{equation}
c_B=\omega^{1/2}\,\left(\frac{EI}{\lambda_m}\right)^{1/4}
\end{equation}
for the first mode of vibration. We can then evaluate the distance the wave has travelled at the end of the impact, i.e. at $t_{po}$. The result of the calculation is presented in Fig. \ref{figure_5}(d) and shows that the pinch-off occurs before the bending waves can travel the entire length of the fiber for length $L\geq 150\,{\rm mm}$. This leads to the oscillations that can be observed in Fig. \ref{figure_5}(c). To characterize the drop capture or release, we consider the bending time before the threshold. In this situation the bending time is well captured by the relation $t_b=\alpha\, \pi/(2\,\omega)$ where $\alpha\simeq1.3$.

\subsection{Effective amplitude of deflection}

As emphasized previously, the drop travels with the fiber during a time $t_{po}$ at which time the drop gets released or recoils. Therefore, the fiber tip travels over an apparent deflection $A_{\rm eff}$ during the drop/fiber interaction time $t_{po}$.

To evaluate the apparent deflection $A_{\rm eff}$, we identify three cases:\\

 (i) when $t_{po} > t_b$, the fiber has enough time to reach its maximum deflection and then starts to oscillate. At the time of the pinch-off, the position of the fiber is
\begin{equation} \label{deflection_A1}
A_{\rm eff}=A\,\sin(\omega\,t_b)+c\,, \qquad {\rm for} \qquad t_{po}>t_b.
\end{equation}\\

(ii) When the length of the fiber is increased, the bending time $t_b$ increases accordingly. For $L \simeq 135\,{\rm mm}$, $t_b \simeq t_{po}$ and the apparent deflection reaches its maximum value $A_{\rm eff} \simeq A+c$.\\

 (iii) Finally, for longer fibers, the fiber tip travels over a distance
\begin{equation} \label{deflection_A2}
A_{\rm eff} \simeq (A+c)\,\frac{t_{po}}{t_b}\, \qquad {\rm for} \qquad t_{po}<t_b.
\end{equation}

Indeed, according to Fig. \ref{figure_5}(a), the fiber travels a distance $A+c$ during a time $t_b$. Therefore, as an approximation the mean velocity of the fiber during its motion is $V_f=(A+c)/t_b$.

\begin{center}
    \begin{figure}
\includegraphics[width=0.47\textwidth]{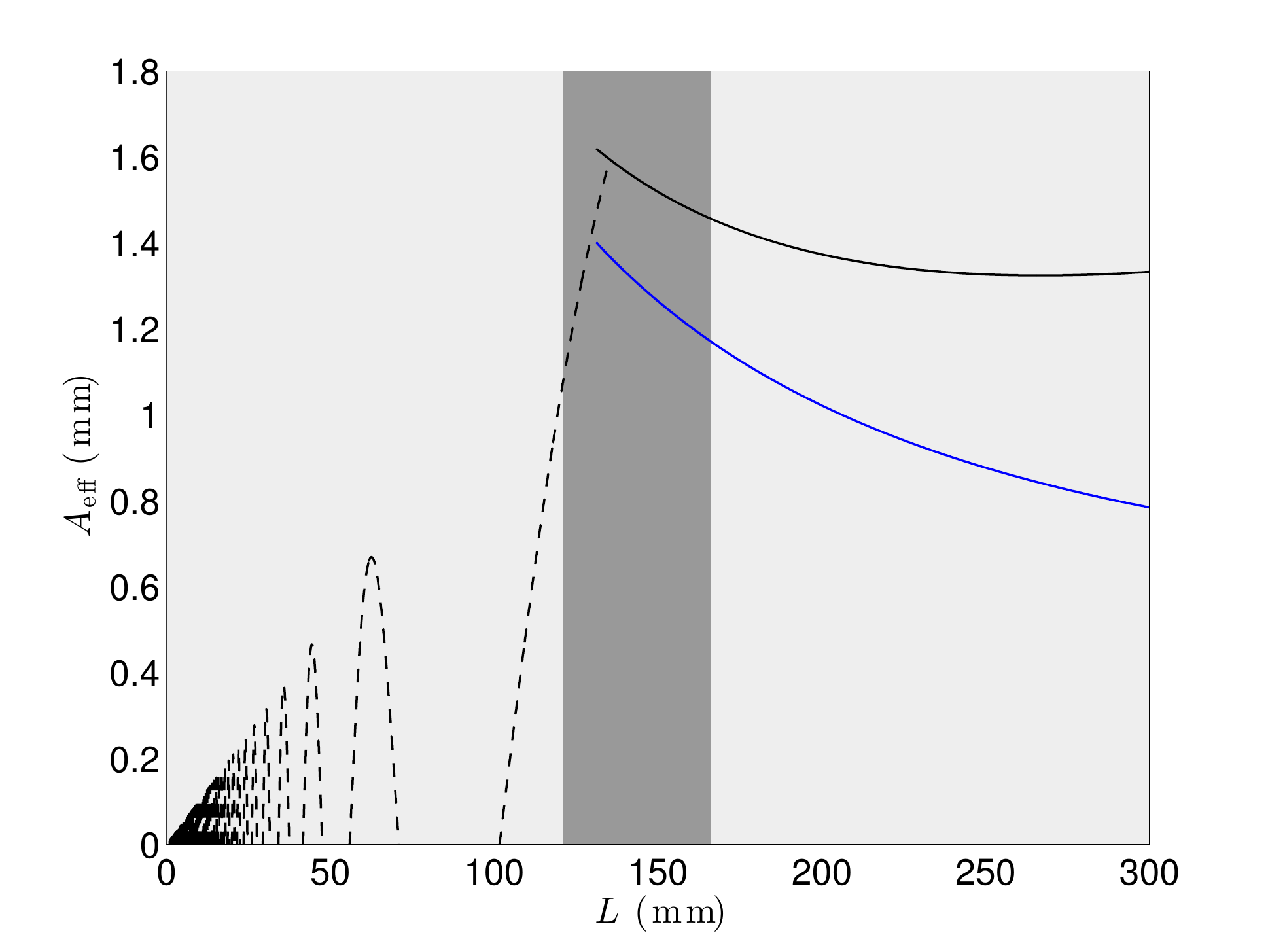}
        \caption{ \small {Qualitative evolution of the effective deflection amplitude $A_{eff}$ of the fiber tip as a function of the length of the fiber plotted for a drop of radius $R=0.76$ mm and a velocity $V_d=450\,{\rm mm.s^{-1}}$. The dotted line represents the position of the fiber at $t_b$ when $t_{b}<t_{po}$, {\it i.e.}, situation (i), and is given by Eq. (\ref{deflection_A1}). For $L > 135\,{\rm mm}$, the situations (ii) $t_b \simeq t_{po}$ and (iii) $t_b > t_{po}$ are plotted using equation (\ref{deflection_A2}) in solid lines. The black and blue solid lines indicate the drop capture and release regimes, respectively.} }
        \label{figure_6}
    \end{figure}
\end{center}

For the pinch-off time measured experimentally, we report in Fig. \ref{figure_6} the expression (\ref{deflection_A2}) for $L > 135\,{\rm mm}$ and the expression (\ref{deflection_A1}) when $L<135\,{\rm mm}$ for both drop capture and release. For small fibers, the dynamics are quite complex since the fiber can oscillate several times before the pinch-off occurs. The maximum value of $A_{\rm eff}$ is reached around $L \simeq 135 \,{\rm mm}$ and corresponds to the maximum deflection $A+c$ reached by the fiber during the drop/fiber interaction time $t_{po}$. Then, as the length of the fiber is further increased the effective amplitude of displacement slightly decreases because the time scale associated with bending of the fiber increases.

\subsection{Threshold velocity}

\begin{center}
    \begin{figure}
\includegraphics[width=0.47\textwidth]{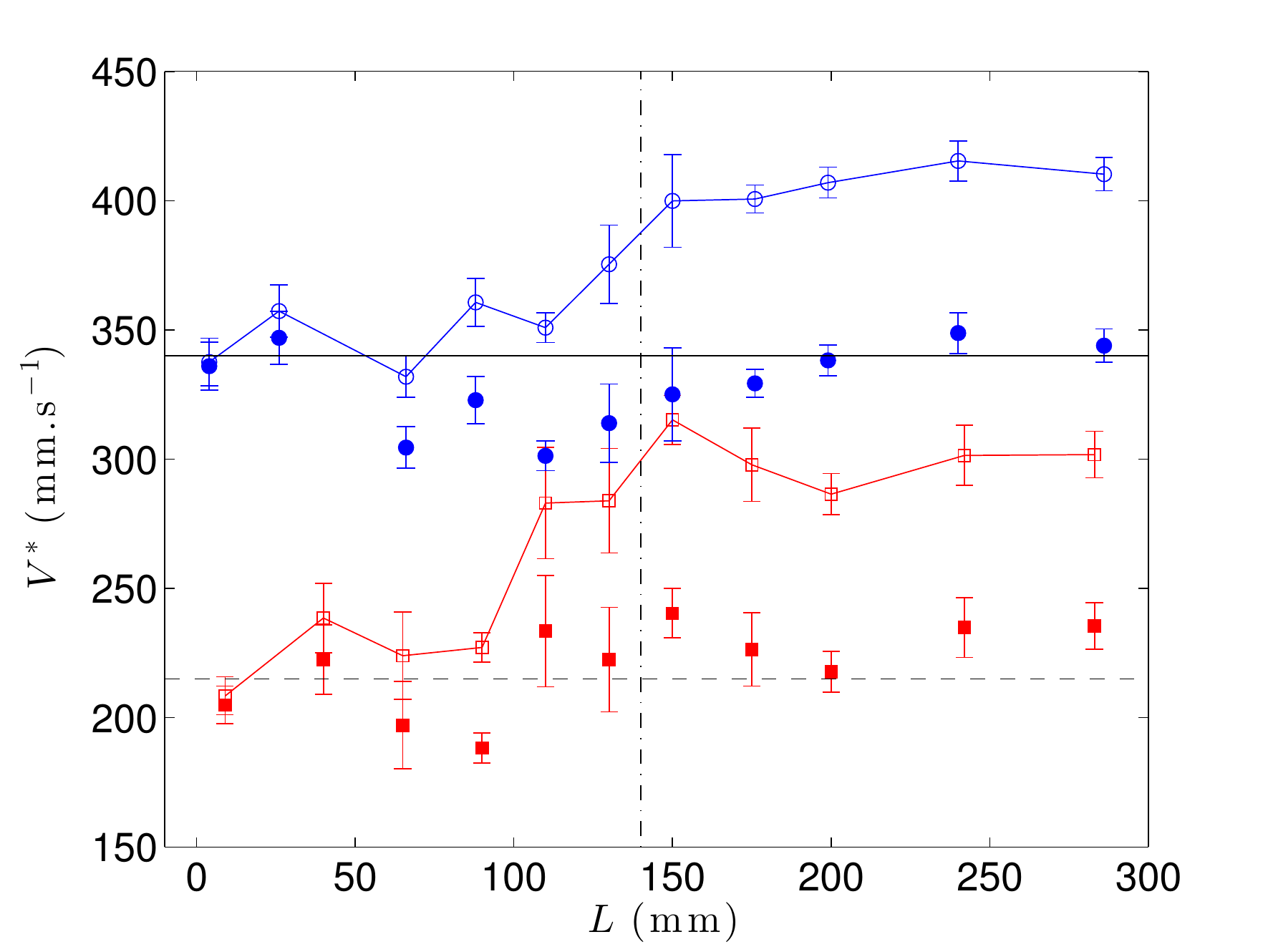}
        \caption{ \small Evolution of the threshold velocity $V_{d}^*$ for capture as a function of the fiber length $L$ for $R=0.76\,\textrm{mm}$ and $\nu=1\,\times 10^{-6}\,{\rm m^2.s^{-1}}$ (red hollow squares) and $\nu=5\,\times 10^{-6}\,{\rm m^2.s^{-1}}$ (blue hollow circles). The threshold velocities are rescaled by the relative speed of the fiber given by equation (\ref{relative_motion}) (filled symbol). The horizontal lines indicate the threshold velocities for $L=0$ mm and the vertical dashed-dotted line corresponds to $L = 140\,{\rm mm}$.}
        \label{figure_7}
    \end{figure}
\end{center}

We can now compare the results of this discussion with the experimental measurement of the threshold velocity for two different viscosities (Fig. \ref{figure_7}). Qualitatively, we observe that the threshold velocity for capture has the same behavior for the two fluids. The threshold velocity $V_d^*$ reaches a maximum for $L \simeq 140\,{\rm mm}$ and then remains roughly constant. The optimal fiber length in our experiments is therefore consistent with the analytical explanation.

When $t_{po} < t_{b}$, i.e. for $L > 140 \,{\rm mm}$, it is useful to consider the moving frame of the fiber tip. This frame is accelerating after the drop impact with an acceleration $-A\,\omega^2\,\sin(\omega\,t)$ leading to a fictitious force. In our experiments, the fictitious force is smaller than inertial effects and negligible for the longer fibers considered here. Assuming that the drop dynamics in the reference frame of the fiber is the same whether the frame is fixed (rigid fiber) or moving (flexible fiber), we can rewrite the threshold velocity as:
\begin{equation}\label{relative_motion}
V_{rel}^*=V_d^*-\frac{A+c}{t_{po}},
\end{equation}
where $V_{rel}^*$ is the threshold velocity in the reference frame of the fiber tip. This indicates that the threshold velocity $V_d^*$ on a flexible fiber can be determined using the threshold velocity on a rigid fixed fiber and the mechanical properties of the flexible fiber (see Fig. \ref{figure_7}). {In summary, using the expression of the threshold obtained by Lorenceau et al. for a rigid fiber \cite{lorenceau2004}, we can estimate $V_{rel}^*$. Then, using the characterization of the response of the fiber following the drop impact presented in the previous section, one can estimate $A$ and $c$ as well as the elongation time $t_{po}$ and therefore obtain the expression of the threshold velocity for a flexible fiber:
\begin{equation}\label{relative_motion}
V_d^*=V_{rel}^*+\frac{A+c}{t_{po}}.
\end{equation}}

\section{Conclusion}

In this study, we have considered the impact of a drop on a flexible fiber clamped at one end. The dynamics of the fiber have been characterized: its vibration frequency, its static displacement and the amplitude of oscillation are predicted by using the beam equation. The drop behavior depends on the impact velocity and two regimes are identified, a capture regime at low impact velocity $V_d$ and a release regime above a threshold velocity for capture $V_d^*$. The interplay between the drop and the fiber involves two time scales: the time for the drop to pinch-off or recoil and the time for the fiber to bend.  The effect of flexibility is maximum when these two time scales are of the same order. If the fiber is too rigid, the amplitude of deflection remains small and the fiber can be considered as nearly rigid. The other rigid-like case occurs when the fiber is too long and its bending time becomes really large compared to the drop elongation time. In this situation, the drop crosses the fiber while the fiber only bends slightly. Finally, we have shown that the influence of the fiber flexibility on the threshold velocity can be quantified with a relative velocity that depends on the fiber deflection, the elongation time of the drop and the threshold velocity on rigid fibers. Although we have illustrated the influence of the fiber flexibility on the threshold velocity and highlighted the importance of the elongation and bending time scales, in more complex situations, the impact can be inclined with respect to the fiber \cite{piroird2009} or can be non-centered \cite{lorenceau2009}. These would lead to more complex fiber and drop dynamics that remain to be investigated.\\
\\

\section*{Appendix A: Influence of the size of the drop}

\begin{center}
    \begin{figure}
    
\includegraphics[width=0.48\textwidth]{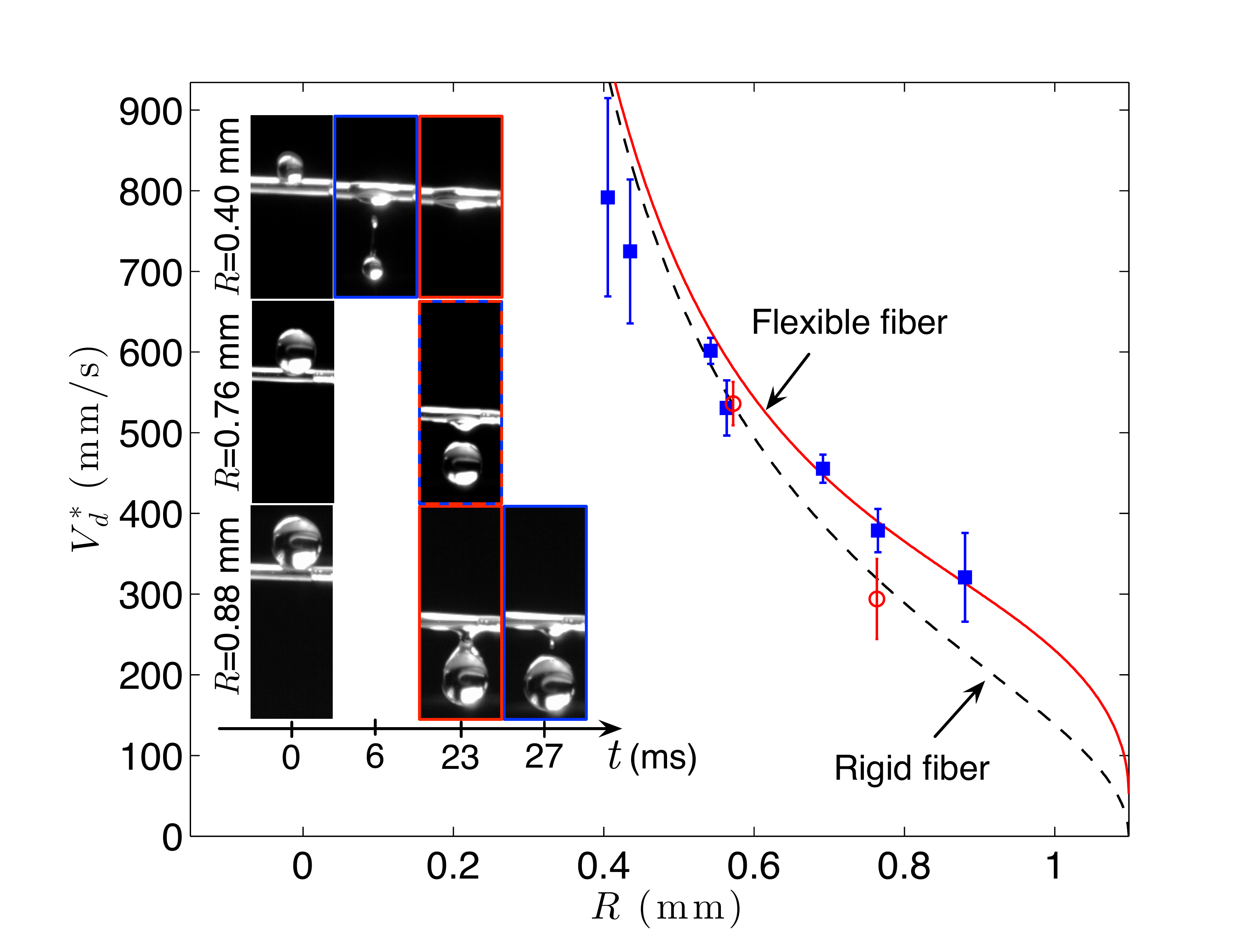}
        \caption{ \small Evolution of the velocity threshold $V_{d}^*$ for capture as a function of the droplet radius $R$ using $\nu=3.5\,\times 10^{-6}\,{\rm m^2.s^{-1}}$  silicone oil and a fiber of length $L=140$ mm (blue square). The red circles indicate the data obtained with a rigid fiber. The black dashed line is the threshold for a rigid fiber obtained using Eq. (\ref{elise}) \cite{lorenceau2004} and the red continuous line is the threshold for a flexible fiber obtained using Eq. (\ref{relative_motion}). The inset shows the time line for three fiber radii highlighting the pinch-off time $t_{po}$ and the time $t_b$ at which the maximum deflection is reached.}
        \label{figure_8}
    \end{figure}
\end{center}

We consider the evolution of the velocity threshold with the drop size. We perform systematic experiments using different needles to obtain a range of droplet radii $R \in [0.43;\,1.04]\,{\rm mm}$. To estimate the influence of the flexibility, we perform these experiments in the situation where we previously found that $t_{po} \sim t_b$ for a drop of radius $R \simeq 0.76$ mm, which corresponds to a fiber length $L \simeq 140\,{\rm mm}$ and $t_{po} \sim t_b \sim 22 \,{\rm ms}$. The results of our investigations are reported in Fig. \ref{figure_8}. We observe that the threshold velocity for capture increases when the size of the drop decreases, which is consistent with the work of Lorenceau et al. on rigid fibers \cite{lorenceau2004}. In particular, Lorenceau et al. showed that the velocity threshold is given by:
\begin{equation}\label{elise}
V_d^*=\Biggl[\frac{{\rm e}^{k\,R_M/R-1}}{k\,R_M/R}\,\left[\left(\frac{R_M}{R}\right)^2-\frac{R}{R_M}\right]\Biggr]^{1/2}
\end{equation}
for a thin rigid fiber. In this expression, $k=12\,a\,C_D/(\pi\,R_M)$, where $C_D$ is the drag coefficient of order unity and $R_M$ is the maximum radius for a drop hanging on a horizontal fiber or radius $a$ with
\begin{equation}
R_M=\left(\frac{3\,a\,\gamma}{\rho\,g}\right)^{1/3}.
\end{equation}

We can now compare the evolution of the threshold velocity for capture as a function of the drop radius. As a first approximation, we consider that the pinch-off time is $t_{po} \sim 22\,{\rm ms}$ (see inset of Fig. \ref{figure_8}). Using the expression $V_{rel}^*=V_d^*-(A+c)/t_{po}$ with $V_d^*$ given by (\ref{elise}) and where $A$ and $c$ depend on the impact velocity and drop mass, we obtain the relative threshold velocity for different drop sizes.

We conclude that for large drops, fiber flexibility leads to a significant increase in the threshold velocity for capture whereas for small drops the effect of the flexibility becomes negligible (Fig. \ref{figure_8}). This qualitative observation is in agreement with the fact that, for small droplets the amplitude of displacement of the fiber becomes smaller whereas the vibration frequency and the bending time of the fiber do not depend on the drop mass. \\

{\footnotesize{ 

\section*{Acknowledgments}

ED is supported by set-up funds by the NYU Polytechnic School of Engineering. FB acknowledges that the research leading to these results partially received funding from the People Programme (Marie Curie Actions) of the European Union's Seventh Framework Programme (FP7/2007-2013) under REA grant agreement 623541. We thank A. Gelperin for sharing his pipette puller for our experiments, E. Lorenceau and E. Villermaux for helpful discussions.\\
}}
    \bibliography{biblio_impact_fiber}
    \bibliographystyle{ieeetr}

    \end{document}